\documentclass{aastex63}
\usepackage{xcolor}
\usepackage{longtable}
\usepackage{amsmath,amssymb}
\usepackage{rotating}

\usepackage{amssymb}
\usepackage{url}

\makeatletter

\newcommand{\Rmnum}[1]{\expandafter\@slowromancap\romannumeral #1@}
\makeatother

\submitjournal{ApJ}

\shorttitle{\textit{Insight-HXMT} observations of A0535+262 in 2020 outburst}
\shortauthors{Wang et al.}

\begin{document}

\title{Timing properties of the X-ray accreting pulsar 1A 0535+262 studied with \textit{Insight-HXMT}}

\author{P. J. Wang\textsuperscript{*}}
\email{wangpj@ihep.ac.cn}
\affil{University of Chinese Academy of Sciences, Chinese Academy of Sciences, Beijing 100049, People's Republic of China}
\affil{Key Laboratory of Particle Astrophysics, Institute of High Energy Physics, Chinese Academy of Sciences, Beijing 100049, People's Republic of China}

\author{L. D. Kong\textsuperscript{*}}
\email{kongld@ihep.ac.cn}
\affil{University of Chinese Academy of Sciences, Chinese Academy of Sciences, Beijing 100049, People's Republic of China}
\affil{Key Laboratory of Particle Astrophysics, Institute of High Energy Physics, Chinese Academy of Sciences, Beijing 100049, People's Republic of China}

\author{S. Zhang}
\affil{Key Laboratory of Particle Astrophysics, Institute of High Energy Physics, Chinese Academy of Sciences, Beijing 100049, People's Republic of China}

\author{V. Doroshenko}
\affil{Institut f\"ur Astronomie und Astrophysik, Kepler Center for Astro and Particle Physics, Eberhard Karls Universit\"at, 72076 T\"ubingen, Germany}
\affil{Space Research Institute of the Russian Academy of Sciences, Profsoyuznaya Str. 84/32, Moscow 117997, Russia}

\author{A. Santangelo}
\affil{Institut f\"ur Astronomie und Astrophysik, Kepler Center for Astro and Particle Physics, Eberhard Karls Universit\"at, 72076 T\"ubingen, Germany}

\author{L. Ji}
\affil{School of Physics and Astronomy, Sun Yat-Sen University, Zhuhai, 519082, People's Republic of China}

\author{E. S. Yorgancioglu}
\affil{Institut f\"ur Astronomie und Astrophysik, Kepler Center for Astro and Particle Physics, Eberhard Karls Universit\"at, 72076 T\"ubingen, Germany}

\author{Y. P. Chen\textsuperscript{*}}
\email{chenyp@ihep.ac.cn}
\affil{Key Laboratory of Particle Astrophysics, Institute of High Energy Physics, Chinese Academy of Sciences, Beijing 100049, People's Republic of China}

\author{S. N. Zhang}
\affil{University of Chinese Academy of Sciences, Chinese Academy of Sciences, Beijing 100049, People's Republic of China}
\affil{Key Laboratory of Particle Astrophysics, Institute of High Energy Physics, Chinese Academy of Sciences, Beijing 100049, People's Republic of China}

\author{J. L. Qu}
\affil{Key Laboratory of Particle Astrophysics, Institute of High Energy Physics, Chinese Academy of Sciences, Beijing 100049, People's Republic of China}

\author{M. Y. Ge}
\affil{Key Laboratory of Particle Astrophysics, Institute of High Energy Physics, Chinese Academy of Sciences, Beijing 100049, People's Republic of China}

\author{J. Li}
\affil{CAS Key Laboratory for Research in Galaxies and Cosmology, Department of Astronomy, University of Science and Technology of China, Hefei 230026, People's Republic of China}
\affil{School of Astronomy and Space Science, University of Science and Technology of China, Hefei 230026,  People's Republic of China}

\author{Z. Chang}
\affil{Key Laboratory of Particle Astrophysics, Institute of High Energy Physics, Chinese Academy of Sciences, Beijing 100049, People's Republic of China}

\author{L. Tao}
\affil{Key Laboratory of Particle Astrophysics, Institute of High Energy Physics, Chinese Academy of Sciences, Beijing 100049, People's Republic of China}

\author{J. Q. Peng}
\affil{University of Chinese Academy of Sciences, Chinese Academy of Sciences, Beijing 100049, People's Republic of China}
\affil{Key Laboratory of Particle Astrophysics, Institute of High Energy Physics, Chinese Academy of Sciences, Beijing 100049, People's Republic of China}

\author{Q. C. Shui}
\affil{University of Chinese Academy of Sciences, Chinese Academy of Sciences, Beijing 100049, People's Republic of China}
\affil{Key Laboratory of Particle Astrophysics, Institute of High Energy Physics, Chinese Academy of Sciences, Beijing 100049, People's Republic of China}

\begin{abstract}
	
	We report results on the timing analysis of the 2020 giant outburst of 1A 0535+262, using broadband data from \textit{Insight-HXMT}. The analysis of the pulse profile evolution from the sub-critical luminosity to super-critical luminosity regime is presented for the first time. 
	We found that the observed pulse profile exhibits a complex dependence on both energy and luminosity.
	A dip structure at the energy of the cyclotron resonant scattering features (CRSFs) is found for the first time in the pulse fraction-energy relation of 1A 0535+262, when the outburst evolves in a luminosity range from 4.8 $\times 10^{37}$ erg s$^{-1}$ to 1.0 $\times 10^{38}$ erg s$^{-1}$.
	The observed structure is luminosity dependent and appears around the source critical luminosity ($\sim$ 6.7 $\times 10^{37}$ erg s$^{-1}$).

\end{abstract}


\section{Introduction}
In X-ray pulsars (XRPs), the accreting material reaches the surface of the neutron star along the magnetic field lines, heating the polar cap, and producing pulsed emission associated with the rotation of the neutron star.
The observed pulse profile shape is largely determined by the configuration of the emission region, which depends on the accretion rate \citep{Basko1976}. 
At low luminosity, the radiation is produced by hot spots or mounds at the polar cap and escapes predominantly along magnetic field lines \citep{Burnard1991,Nelson1993}, which results in the so-called ``pencil beam''.
At high luminosity, the radiation pressure is sufficient to effectively decelerate the accreted plasma, which results in the formation of a radiatively-dominated shock above the polar cap, and of an extended emission region, the so-called accretion column. In this case, the radiation mainly escapes through the column walls forming a ``fan beam'' \citep{Basko1976,Wang1981}.
The transition between the low-luminosity and the high-luminosity regimes occurs at the critical luminosity, in some cases determined by studying the correlation between the energy of cyclotron resonant scattering features (CRSFs) and luminosity \citep{Vasco2011,Becker2012,Doroshenko2017,Kong2021}. At sub-critical luminosities, the correlation remains flat or positive, while at super-critical luminosities, the CRSFs-energy is negatively correlated with luminosity \citep{Tsygankov2006,Staubert2019}.  

The change of the emission pattern may result in a significant change in the pulse profile \citep{Doroshenko2020,Ji2020}. The pulse profile of some sources (e.g. 1A 0535+262, 4U 0115+63, Her X-1) is known to exhibit a complex evolution with energy and luminosity, and in particular around the expected critical luminosity. It has therefore been suggested that strong variations around the critical luminosity is associated with a change in the geometry of the emission region \citep{Bulik2003,Annala2010}. 
Although some models based on the analysis of pulse profiles attempt to restore the geometry of the accretion, the real pattern and the geometry of the emitting region remain difficult to study \citep{Lutovinov2009,Caballero2011}. 
In some sources (e.g. V0332+53, 4U 0115+63), pulse profile are also observed to change around the cyclotron line energy \citep{Tsygankov2006,Lutovinov2009}. In particular, the increase/decrease of the depth of the line is expected to also affect the pulse fraction, a measure of the amplitude of the pulsations. The energy dependence of the pulse fraction breaks away from the smooth trend and presents a local hump or dip \citep{Lutovinov2009,Tsygankov2010}.

We note that not all XRPs exhibit significant changes of the pulse profile shape/pulse fraction around the cyclotron line energy, and generally speaking, the relation between the observed pulse profile shape, cyclotron line properties, and emission region geometry are not trivial. However, in the case when changes (of CRSF properties, pulse profiles, accretion regime transition) are observed, they typically change coherently. As of now, there are no sound theoretical models describing such variations. Nevertheless, studying these variations is essential from a phenomenological point of view, and constitutes a key input to refine theoretical models. Such a study is the main goal of the current investigation. To this avail, we firstly report the local sharp reduction of the pulse fraction-energy relation near the CRSFs-energy through this 2020 giant outburst with high-cadence and broad-band observations for 1A 0535+262.

1A 0535+262 is a transient X-ray accreting pulsar, discovered by Ariel V during the giant outburst in 1975 \citep{Rosenberg1975,Coe1975}. The pulsation period and distance are measured to be 103 s and 2 kpc, respectively \citep{Camero-Arranz2012,Bailer-Jones2018}. Its orbital period and eccentricity are 111 days and 0.47, respectively (\cite{Finger1996} and references therein). The optical companion is identified as O9.7 IIIe star HDE 245770 \citep{Giangrande1980}.
Since 1975, 1A 0535+262 is observed to regularly exhibit both normal (Type I) and giant (Type II) outbursts. The type I (normal) outbursts occur with a lower peak luminosity ( $\leq$
${10}^{37}$ erg s$^{-1}$) compared to the type II (giant) outbursts and are typically associated with periastron passages when the neutron star comes closest to the circumbinary disk of the companion.
Observations of normal outbursts have focused on multiple outbursts in 2005, 2007, 2009, and 2010. Although the peak luminosity of the outbursts vary, the pulse profiles at similar luminosities and energy are similar, except for a pre-outburst five days before the 2005 August/September outburst, which is thought to be caused by magnetic field instabilities \citep{Caballero2008}.
The type II giant outburst luminosity usually covers the type I outburst luminosity, up to a few times of the Eddington luminosity at its maximum. There have been about 10 giant outbursts in the history of 1A 0535+262: in 1975, 1977, 1980, 1983, 1989, 1994, 2005, 2009, and 2011, but with relatively sparse investigations (See \cite{Camero-Arranz2012} and references therein). 

In the 2020 outburst, the peak luminosity reached around 1.2 $\times 10^{38}$ erg s$^{-1}$, exceeding, for the first time, the critical luminosity reported by \cite{Kong2021} based on analysis of \textit{Insight-HXMT} observations.
Here we present the timing analysis results based on the same excellent dataset.
In particular, we focus on the analysis of the evolution of the pulse profile morphology, the dependence of the pulse profile with energy, and the possible influence of CRSFs on the dependence of the pulse fraction on energy and luminosity.
The observations and data processing are given in Section \ref{obs}, and the results are presented in Section \ref{result}. In Section \ref{diss}, we compare the result of the 2020 outburst with previous observations and other accreting pulsars.

\section{Observations and Data analysis}
\label{obs}
\textit{Insight-HXMT} is China's first X-ray astronomy satellite, launched in 2017. Since its launch, \textit{Insight-HXMT} has performed multiple high-cadence observational campaigns for a number of bright transients, which has led to a large number of interesting results \citep{Doroshenko2020,Kong2020,Ma2021,You2021}. The scientific payload of \textit{Insight-HXMT} comprises three main detectors: the High Energy X-ray Telescope (HE, poshwich NaI/CsI, 20--250 keV, \cite{Liu2020,Zhang2020}), the Medium Energy X-ray Telescope (ME, Si-PIN detector, 5--40 keV, \cite{Cao2020}) and the Low Energy X-ray telescope (LE, SCD detector, 1--10 keV, \cite{Chen2020}).

\textit{Insight-HXMT} performed an extensive observational campaign of the 2020 outburst of 1A 0535+262, with a total exposure time about 1.910 Ms, covering almost the entire outburst period. We extracted the event data using the \textit{Insight-HXMT} data analysis software {\tt HXMTDAS v2.04} with standard \textit{Insight-HXMT} Data Reduction Guide {\tt v2.04}\footnote[1]{{http://http://hxmtweb.ihep.ac.cn/SoftDoc.jhtml}}, discarding the data of detector boxes affected by the Crab \citep{Kong2021}.
Then, we extracted the net light curve for folding the pulse profile.
The sky map of the FoVs of each detector and the location of the contaminant sources are presented in detail in \cite{Kong2021}. 
The arrival times of photons was corrected to the Solar system barycentre with the {\tt HXMTDAS} tool {\tt hxbary}, and the effects of binary orbital modulation is corrected by the ephemerides provided by \cite{Camero-Arranz2012}.
For each observation, we obtained the pulse period using an epoch-folding method.
We selected the 70--90 keV pulse profile at MJD 59167 as the template to calculate the deviation of the phase of the pulse profile at same energy for each observation, using the {\tt fftfit} function \citep{Taylor1992} in {\tt stingray.pulse} \citep{Huppenkothen2019}.
Then, for all energy ranges, the pulse profiles of each observation were aligned with each other by the deviation of the phase and were arranged as a function of time.

\section{Results}
\label{result}
\subsection{Luminosity-dependence of the pulse profile}
\label{pulse profile evo}

\begin{figure}
	\centering
	\includegraphics[angle=0,scale=0.45] {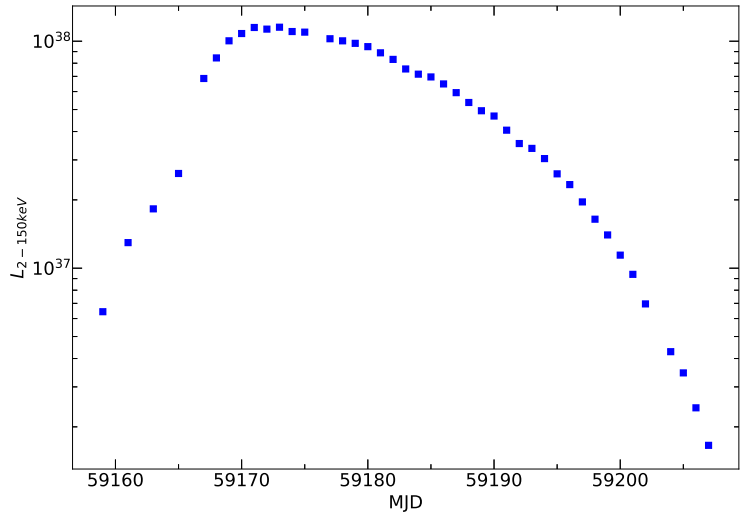}
	\caption{The light curve of 2020 giant outburst. Each point represents the luminosity of one day.
	}
	\label{lum}
\end{figure}

\begin{figure*}
	\centering
	\includegraphics[angle=0,scale=0.35] {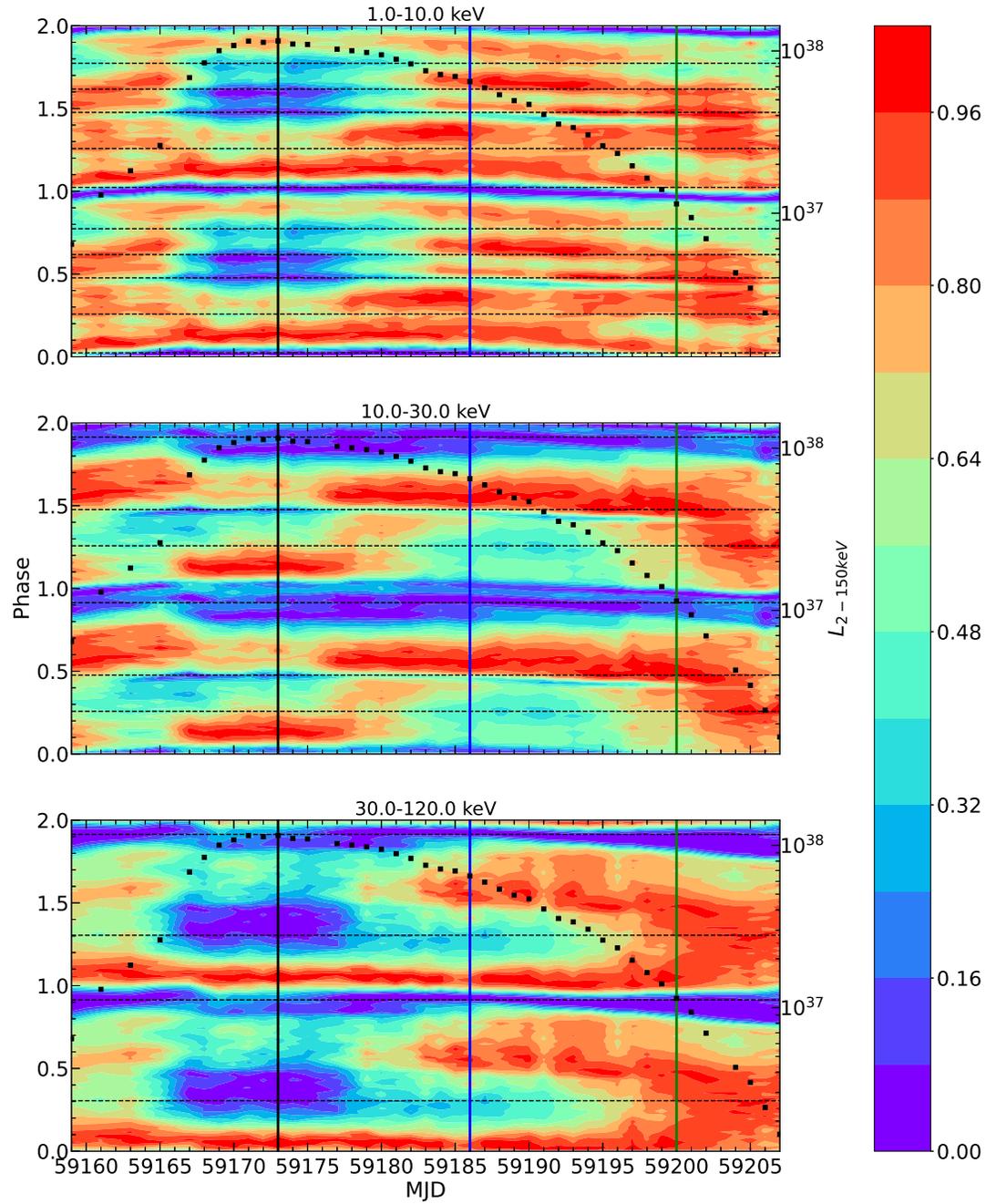}
	\caption{Pulse profiles vs. time expressed in MJDs; the top, middle, and bottom panels show the evolution of the pulse profiles in the energy ranges 1.0--10.0 keV, 10.0--30.0 keV, and 30.0--120.0 keV, respectively. The color bar displays the intensity of the pulse profile normalized in the [0,1] range.
	The black solid line at MJD 59173 and blue solid line at MJD 59186 indicate the peak luminosity and the critical luminosity of the outburst, respectively. The green solid line indicates the position at MJD 59200. The black dashed lines roughly indicate the position of the dips observed in the pulse profiles.
	}
	\label{pro_evo_zong}
\end{figure*}

\begin{figure*}
	\centering
	\hspace{-0.4cm}
	\vspace{-0.0cm}
	\includegraphics[angle=0,scale=0.37] {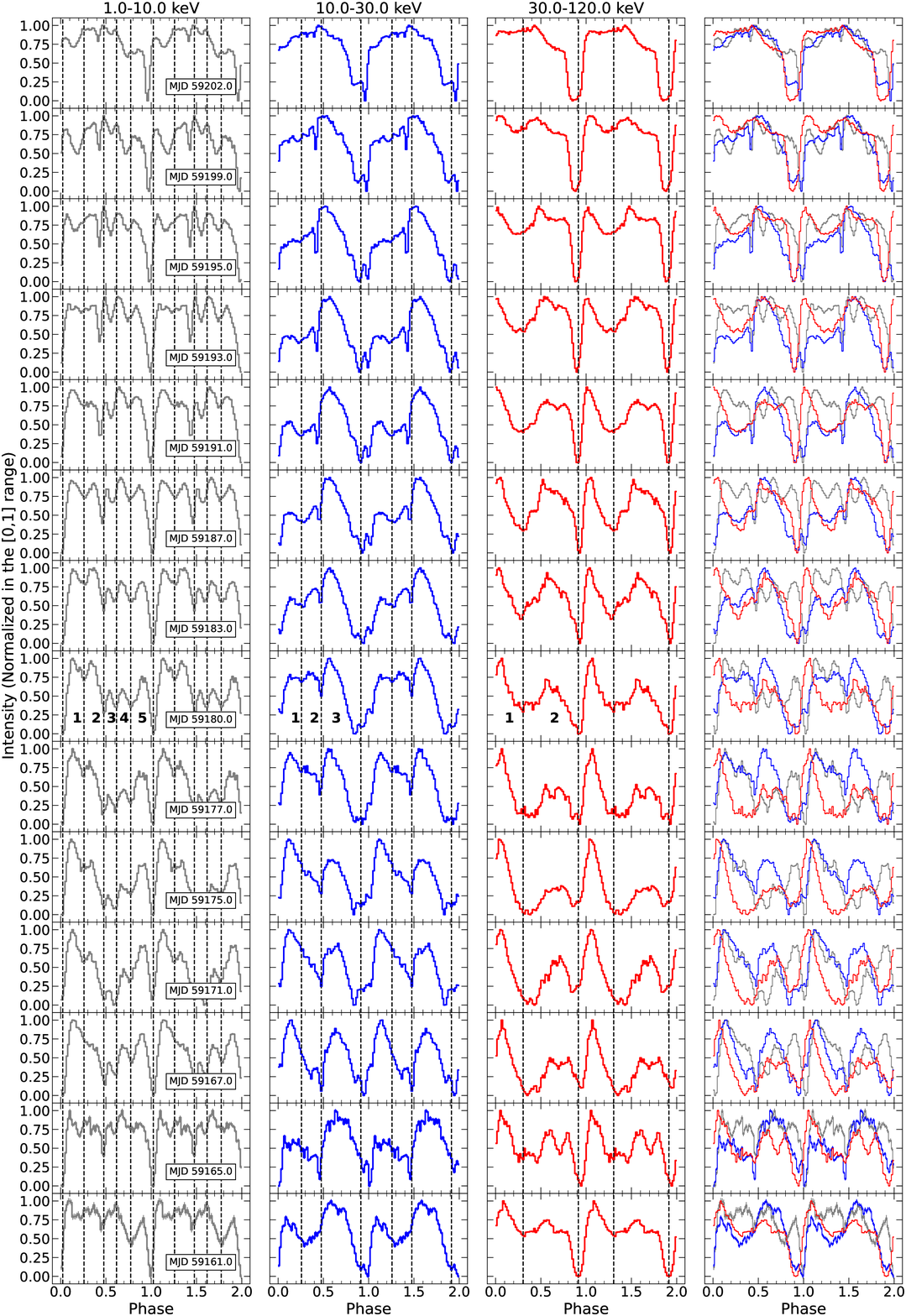}
	\caption{
		Panel 1 (From left to right): The gray lines show examples of pulse profiles with different observing times in the range 1.0--10.0 keV. Panel 2: The observing time of each sub-panel is the same of that in panel 1, but in the energy range 10.0--30.0 keV. Panel 3: Same as panel 2, but for 30.0--120.0 keV. Panel 4: The gray lines, blue lines, and red lines show pulse profiles in the energy ranges 1.0--10.0 keV, 10.0--30.0 keV, and 30.0--120.0 keV.
		The intensity of the pulse profile is normalized in the [0,1] range.
		The black dashed lines are the same as in Figure \ref{pro_evo_zong}. The error bars of the pulse profiles are smaller than the points, which is hard to be recognized in the figure.
	}
	\label{pro_evo_one}
\end{figure*}

In general, the pulse profile of an accreting pulsar evolves with luminosity.
Based on high-cadence observations by \textit{Insight-HXMT}, we obtain an almost complete evolution of the pulse profiles throughout the 2020 outburst.
The luminosity of the 2020 outburst versus time is shown in Figure \ref{lum}. The outburst reaches its maximum luminosity at MJD 59173. The duration of the rising phase of the outburst is about 15 days, lasting from MJD 59158 to MJD 59173, shorter than the duration of the fading phase. The luminosity rises sharply during the rising phase, with a large span of luminosity between each observation. For the fading phase between MJD59173 and MJD59207, \textit{Insight-HXMT} observations are denser, covering a luminosity range of 1.7 $\times 10^{36}$ erg s$^{-1}$ to 1.2 $\times 10^{38}$ erg s$^{-1}$.
The pulse profiles for the three \textit{Insight-HXMT} instruments are presented in Figure \ref{pro_evo_zong}.
Pulse profiles in the three energy ranges show abrupt changes around MJD 59166. These are due to the large observational luminosity gap between MJD 59165 and 59167
in the rising phase of \textit{Insight-HXMT} monitoring (See Figure \ref{lum}). We therefore focus on the analysis of the evolution of pulse profiles in the fading period, with more high-cadence observations and a wider luminosity span. During this phase, the evolution of the pulse profile is relatively smooth. However, the luminosity-dependence of the pulse profiles can be discerned in Figure \ref{pro_evo_zong}. The blue line in Figure \ref{pro_evo_zong} corresponds to the source reaching the critical luminosity during the fading phase \citep{Kong2021}.

The pulse profiles in the soft band (1.0--10.0 keV) exhibit the most complex structure: five peaks can be identified up to around the critical luminosity transition (MJD 59186). This behaviour is consistent with previous investigations of the source \citep{Bradt1976,Caballero2007}. However, before the 2020 monitoring campaign of \textit{Insight-HXMT}, no broad-band observations at higher luminosities were available. Some changes are revealed by \textit{Insight-HXMT} observations; the pulse profile gets smoother overall above the critical luminosity.

Compared to those in the energy range 1.0--10.0 keV, the pulse profiles at 10.0--30.0 keV
appear to be overall simpler, and exhibit only three peaks up to the critical luminosity. Again, this is consistent with past investigations of the source \citep{Camero-Arranz2012}. At higher luminosities, \textit{Insight-HXMT} reveals gradual changes illustrated in the middle panel of Figure \ref{pro_evo_zong}. In particular, 
the structure of peak 2 disappears and the intensity of peak 1 is gradually higher than that of peak 3. When the luminosity decreases below the critical luminosity, the dip between peak 2 and peak 3 gradually disappears. The pulse profile evolves from double peaks to a single peak. 

At energies above 30.0 keV, the detailed results on the evolution of the profile with luminosity and time are presented in the bottom panel of Figure \ref{pro_evo_zong} and the panel 3 of Figure \ref{pro_evo_one}. The pulse profile switches between single and double peaks around MJD 59200, corresponding to a luminosity of 1.1 $\times 10^{37}$ erg/s. This luminosity also corresponds to the location of pulse profile transitions for 10.0--30.0 keV, and is somewhat lower than the critical luminosity value $\sim$ 6.7 $\times 10^{37}$ erg/s derived by \cite{Kong2021} based on analysis of the CRSFs evolution with luminosity. Nevertheless,
these changes, which are likely associated with a strong change of the emission pattern of the pulsar, are close to the observed critical luminosity. This suggests that it could be associated with the transition from the super to sub-Eddington regime.
At luminosities below 1.1 $\times 10^{37}$ erg/s, the pulse profile shows a single peak, similar to that of quiescence, which is most likely related to a pencil beam. 


Overall, the high-cadence \textit{Insight-HXMT} observations of the fading phase of the 2020 outburst are consistent with observations taken at a similar luminosity during previous normal and giant outbursts. However, the maximum luminosity observed during the 2020 monitoring campaign is 1.2 $\times 10^{38}$ erg s$^{-1}$, higher than the critical luminosity by a factor of $\sim$1.73. As a consequence, 1A 0535+262 is the second source robustly observed in the super-critical luminosity regime \citep{Kong2021}. 
The observed pulse profile shape changes discussed above appear to support this conclusion. It is worth, however, to take a more detailed look at the energy dependence of the pulse profiles at each luminosity, in particular around the cyclotron line energy to link the two investigations.

\subsection{Energy dependence of the pulse profile}
\label{pulse profile vs. energy}
\begin{figure*}
	\centering
	\includegraphics[angle=0,scale=0.4] {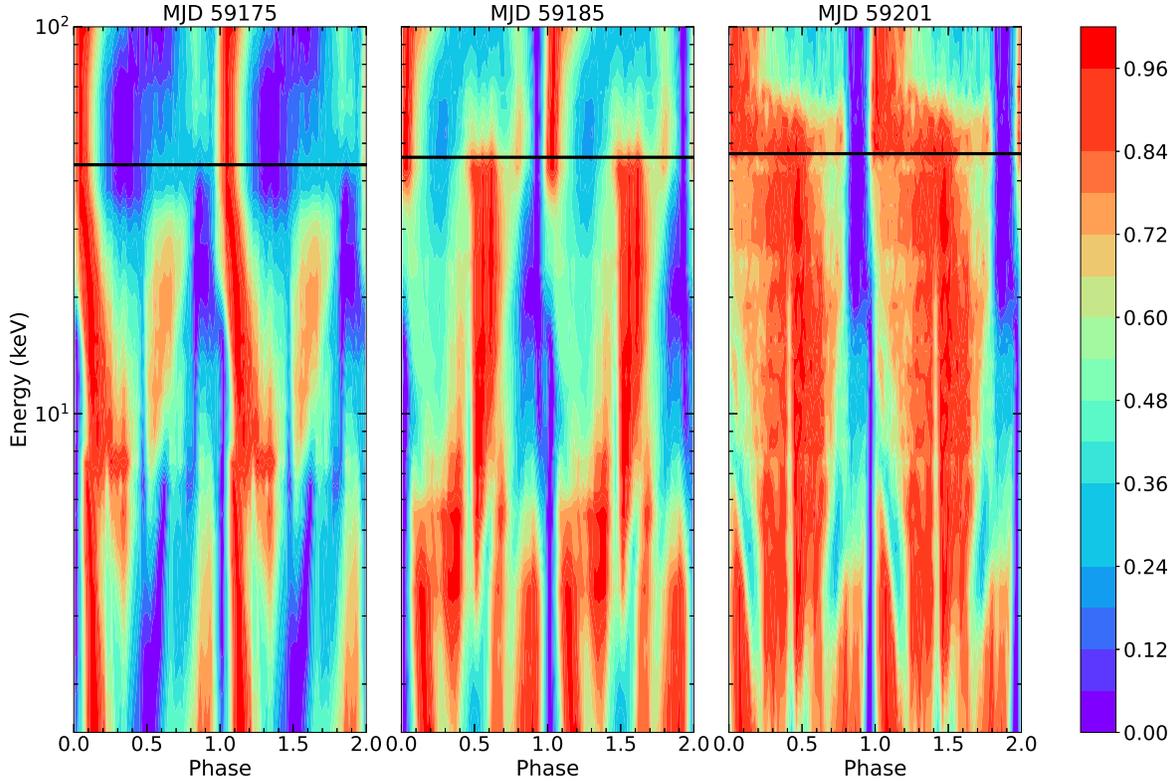}
	\caption{Pulse profiles vs. energy. The three panels (From left to right) show observations at MJD 59175 (around the peak luminosity), MJD 59185 (around the critical luminosity) and MJD 59201 (at low luminosity).
	The color bar displays the intensity of the pulse profile normalized in the [0,1] range. The black line indicates the energy of the fundamental CRSFs of each observation time.
	}
	\label{pro_E_zong}
\end{figure*}

\begin{figure*}
	\centering
	\hspace{-0.4cm}
	\vspace{-0.0cm}
	\includegraphics[angle=0,scale=0.38] {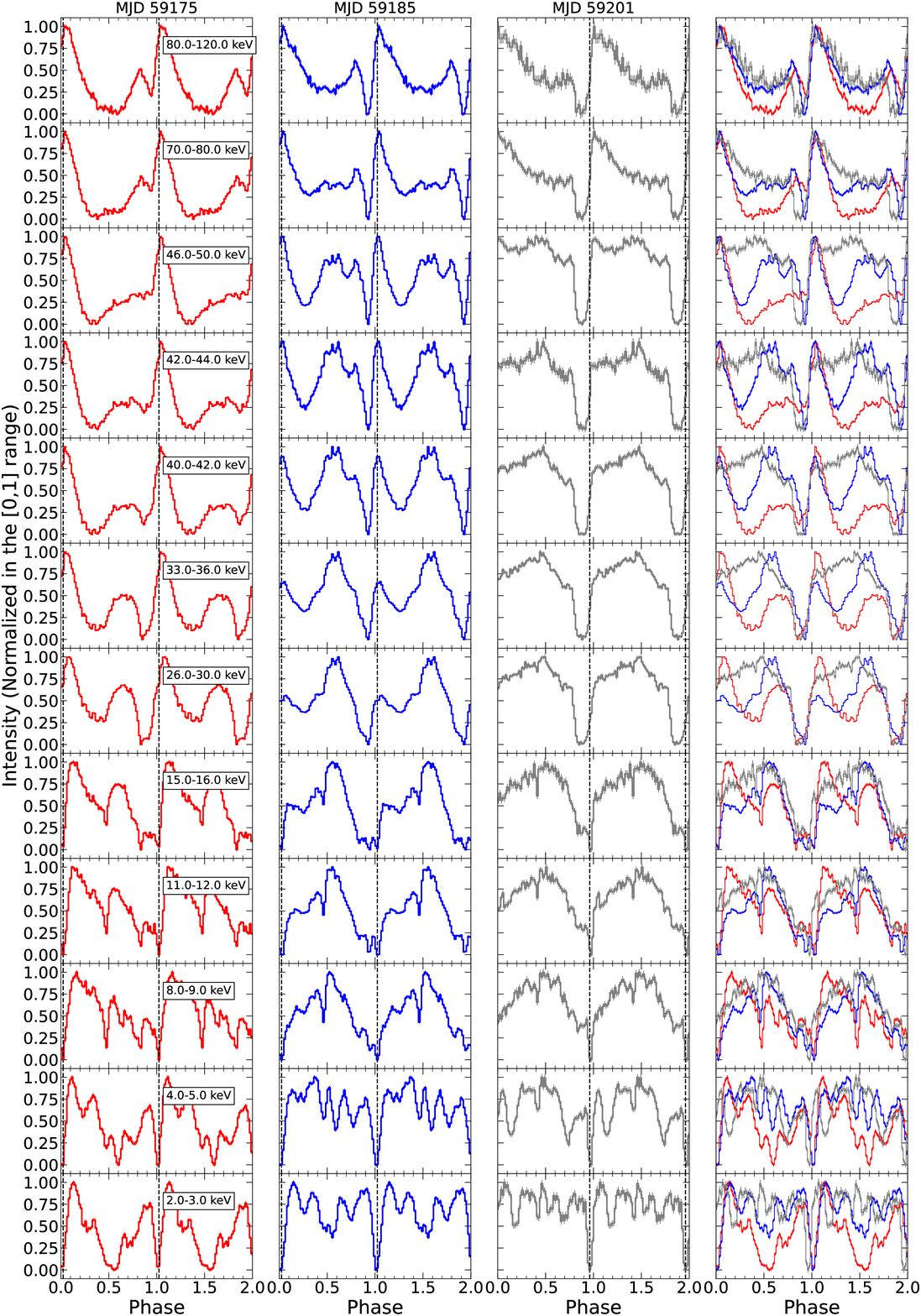}
	\caption{Examples of pulse profiles vs. energy; Panel 1, panel 2, and panel 3 (From left to right) show pulse profiles at MJD 59175, MJD 59185 and MJD 59201. Panel 4: The red lines, blue lines and gray lines indicate the pulse profile in panel 1, panel 2, and panel 3, respectively. 
	The intensity of the pulse profile is normalized in the [0,1] range.
	The black dash lines in each panel mark the global minimum phase in the 2.0--3.0 keV energy range.}
	\label{pro_E_one}
\end{figure*}

The high-statistic and high-cadence observations of \textit{Insight-HXMT} allow us to investigate the energy dependence of the pulse profile.
As mentioned in the previous section, the pulse profiles exhibit different shapes at different luminosities and time.
We therefore investigate the three regimes separately by performing detailed analysis for three representative observations.

The observation described in panel 1 of Figure \ref{pro_E_zong} correspond to a luminosity of 1.1 $\times 10^{38}$ erg s$^{-1}$, close to the peak luminosity of the 2020 outburst. The evolution of the peak phase and dip phase of the pulse profile can be obtained from the two-dimensional diagram. 
As the energy increases, the peak position changes from phase $\sim$ 0.1--0.2 at the lowest energies to $\sim$ 0.0--0.1 at energies above 20 keV. The pulse profiles as a function of energy are shown in panel 1 of Figure \ref{pro_E_one}. 
Above 10 keV, the peak phase of the second peak (0.5--0.7 phase) presents a weak drift with increasing energy, with an offset of about 0.2 phase (Panel 1 of Figure \ref{pro_E_zong}). 
A similar behaviour is, however, not observed for observations at lower luminosities, and in particular, around the transition luminosity.
In the vicinity of the critical luminosity, the peak at 0.5 phase (corresponding to peak 3 in panel 2 of Figure \ref{pro_evo_one}) exhibits no evolution at lower energies, but jumps abruptly at 0.8 phase, at energies larger than the CRSFs energy ($\sim$ 46 keV).
It is noted that the pulse profile changes dramatically at both ends of the CRSFs energy as the luminosity approaches or rises above the critical luminosity, as already observed for V0332+53 and 4U 0115+63 (see detailed discussion below).

At even lower luminosities (\textless 1.1 $\times 10^{37}$ erg s$^{-1}$), away from the critical luminosity, the pulse profile below 6 keV shown in Figure \ref{pro_E_one} is as complex as the one at the high luminosity. However, above 6 keV, the pulse profiles show a single broad peak. The pulse profiles are clearly different from the high luminosity case shown in panel 1 and 2 of Figure \ref{pro_E_one}. The dashed lines in panels 1, 2, and 3 of Figure \ref{pro_E_one} represent the global minimum phase in 2.0--3.0 keV. It is seen that the phase corresponding to the dip of the pulse profile at low energy (\textless 10.0 keV) is the location of the peak at high energy (\textgreater 30.0 keV), for the luminosity near or above the critical luminosity; the origin of this phenomenon may be related to the formation of the accretion column.

\subsection{Energy-Luminosity dependence of the pulsed fraction}
\label{Pulse fraction spectrum}

\begin{figure*}
	\centering
	\includegraphics[angle=0,scale=0.5] {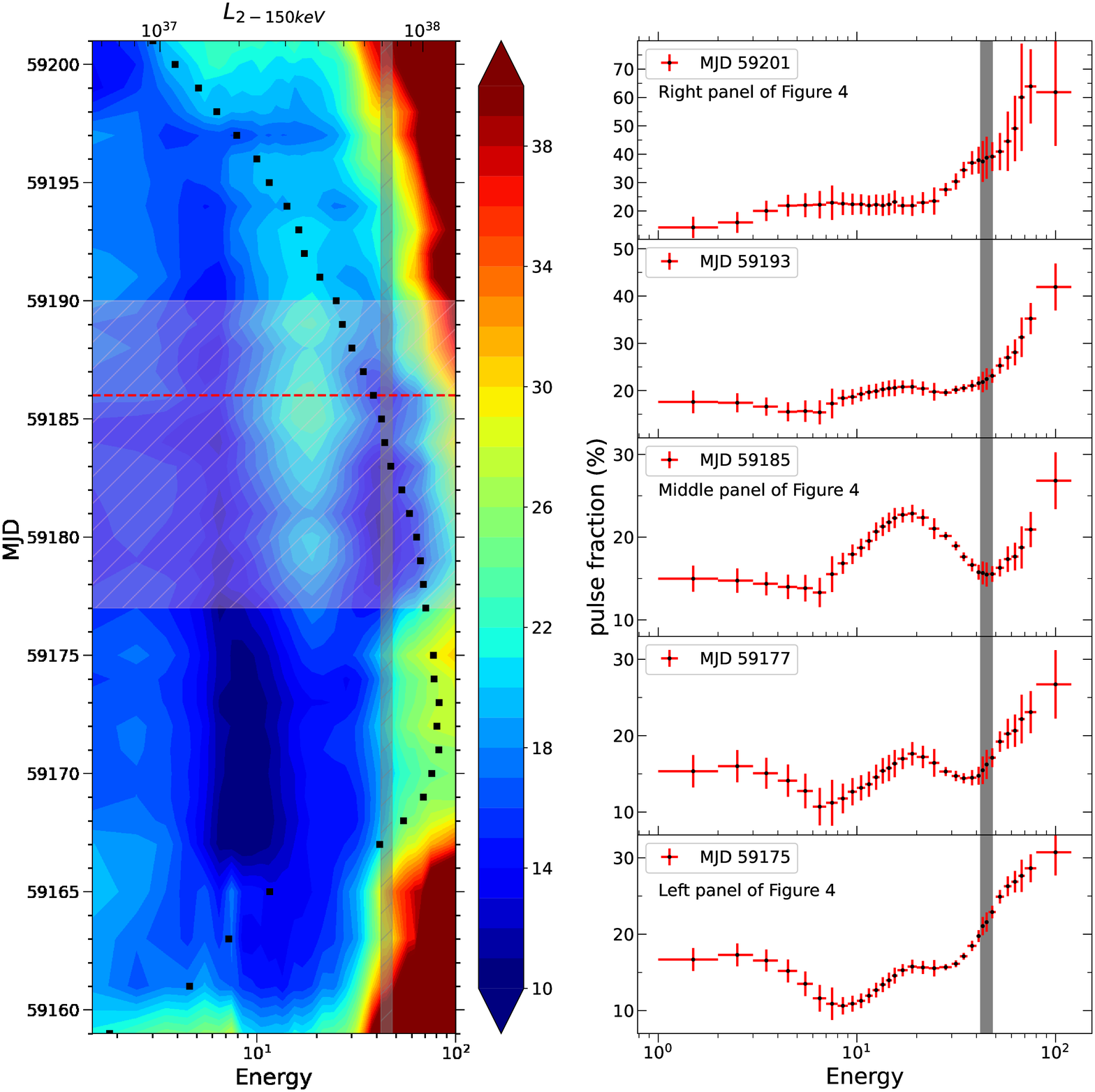}
	\caption{
		Left panel: RMS pulse fraction-energy relation vs. MJD. The red dashed line marks the critical luminosity. The pink shaded area indicates the region where the RMS pulse fraction energy relation has a local sink structure near the CRSFs-energy. Right panel: The typical RMS pulse fraction spectrum at different times. The gray shaded area indicates the CRSFs-energy.
	}
	\label{pf_spec_evo}
\end{figure*}
The amplitude of the pulsations as a function of luminosity and energy can be used to characterize the observed changes in a more quantitative way. We use the RMS pulse fraction to quantify the pulse profile. The RMS pulse fraction is defined as follows \citep{Wilson-Hodge2018,Wang2020}:

\begin{equation}
	{f}_{\rm rms} = \frac{(\sum\nolimits_{i=1}^N(r_i-\bar{r})^2 /N)^{1/2}}{\bar{r}}
\end{equation}
where $\bar{r}$ and $r_i$ are the phase average count rate and the phase count rate, respectively. $N = 64$ is the total phase bin number. The error of ${f}_{\rm rms}$ is estimated by error propagation, and quoted at the 90\% confidence level.

To investigate the energy and luminosity dependence of the pulsed fraction, we calculate ${f}_{\rm rms}$ as a function of energy for each observation. We present results in the left panel of Figure \ref{pf_spec_evo}. For reference, we also separately plot slices corresponding to observations discussed in Section \ref{pulse profile vs. energy} in the right panel. As one can see in Figure \ref{pf_spec_evo}, at MJD 59175, where the luminosity is high, the RMS pulse fractions show a decreasing trend in the low-energy band. 
Above 10 keV, the RMS pulse fraction gradually rises as expected for accreting pulsars.
At MJD 59185, the pulse fraction-energy relation becomes more complex, and exhibits a sharp decrease around the energy of the CRSFs. ($\sim$ 44 keV). 
As seen in Figure \ref{pf_spec_evo}, this phenomenon is only present within MJD 59177--MJD 59190, at luminosities around the critical-luminosity from 1.0 $\times 10^{38}$ erg s$^{-1}$ to 4.8 $\times 10^{37}$ erg s$^{-1}$. The depth of the pulse fraction dip reaches the maximum near the critical luminosity, and then gradually disappears as the luminosity rises. The RMS pulse fraction dip has a large width, and so the evolution of its centroid energy is hard to be investigated. However, it appears to be correlated with the observed line energy.
We also compare the traditional pulse fractions (($(F_{\rm max}-F_{\rm min})/(F_{\rm max}+F_{\rm min})$), the results obtained are similar to those obtained by the RMS pulse fraction \citep{Mandal2022}.
As the luminosity drops below 4.8 $\times 10^{37}$ erg s$^{-1}$ (See the RMS pulse fraction spectrum of MJD 59193 and MJD 59201 in Figure \ref{pf_spec_evo}), the RMS pulse fraction-energy dependence again simplifies, with the RMS pulse fraction rising gradually with energy between 1.0--120.0 keV, similar to that generally observed in accreting pulsar systems.

\section{Discussion}
\label{diss}

We have investigated the brightest ever observed outburst of 1A 0535+262 detected by \textit{Insight-HXMT} in 2020, which monitored the source between 1.7 $\times 10^{36}$ erg s$^{-1}$ and 1.2 $\times 10^{38}$ erg s$^{-1}$. As discussed in \citep{Kong2021}, the relation between the luminosity and the CRSFs-energy allows us to conclude that this outburst reached the super-critical luminosity regime.
The broad-energy and high-cadence of \textit{Insight-HXMT} observations have provided the opportunity to study in detail the source behavior around the critical luminosity, monitoring the transition from the sub-critical to super-critical accretion regime.
For the first time, we were able to investigate the evolution of the pulse profile shape both with energy and luminosity in a very broad energy/luminosity range. The only comparable investigation in the literature has been published by \cite{Lutovinov2015} for V0332+53, however, even in this case no detailed analysis of the pulse profile dependence on energy was reported. On the other hand, statistics of \textit{Insight-HXMT} observations allowed for discerning fine structure in the energy dependence of the pulse fraction associated with the cyclotron line for the first time. 

The transition from the sub-critical to super-critical accretion regime is expected to be accompanied by changes in the geometry of the emission region, which in turn changes the pulse profile shape.
These changes are indeed observed. However, the transition luminosity derived from the CRSFs behaviour by \cite{Kong2021} does not fully coincide with that estimated in our analysis based on the profile morphology.
The observed strong changes of the pulse profile shape occur at luminosity (1.1 $\times 10^{37}$ erg/s) which is close to, but does not exactly, correspond to the critical luminosity value estimated by \cite{Kong2021} (6.7 $\times 10^{37}$ erg/s).


The average spectrum parameters (e.g. Photon index, exponential folding energy, and the temperatures of the blackbody
components) remains stable at low-luminosity, but varies dramatically at high-luminosity, appears to undergo a transition from a flat to drastic evolution, near 1.1 $\times 10^{37}$ erg/s (Figure 5 in \cite{Kong2021}).
The position of the transition revealed by spectrum corresponds to the luminosity of the changes of pulse profiles; but in the vicinity of 6.7 $\times 10^{37}$ erg/s, both the pulse profiles and the spectrum (except the CRSFs line) evolve smoothly \citep{Kong2021}. Those phenomena seem to support the idea that these are two distinct transitions occurred at 1.1 $\times 10^{37}$ erg/s and 6.7 $\times 10^{37}$ erg/s, respectively. However, it can not be ruled out that the two phenomena are due to the same regime transition which, however, is not instantaneous but occurs within a certain luminosity range.
In this regard, one could view the observed pulse profile shape changes as a pre-cursor for the transition revealed by the change between the positive/negative correlation of the CRSFs line energy with luminosity.
The change of the pulse profile morphology occurs at lower flux, and the pulse profile evolves smoothly around the critical luminosity value estimated by \cite{Kong2021} (Figure \ref{pro_evo_zong}). 

Above the critical luminosity ($\sim$ 6.7 $\times 10^{37}$ erg s$^{-1}$), as shown in middle panel of Figure \ref{pro_evo_zong}, the pulse profile remains stable until the luminosity reaches 9.5 $\times 10^{37}$ erg s$^{-1}$ (at MJD 59180), then the maximal pulse phase jump from $\sim$ 0.6 to $\sim$ 0.2 (10.0--30.0 keV); which mean, the intensity of peak 1 becomes higher than peak 3 (Figure \ref{pro_evo_one}).
The phenomenon of a delayed pulse profile change was also hinted in observations of V0332+53, but with no detailed analysis reported in literature \citep{Tsygankov2006}. V0332+53 is the only source where the transition from sub- to super-critical accretion has been unambiguously observed previously \citep{Doroshenko2017,Vybornov2018,Staubert2019}.
No detailed analysis of the pulse profile shape was possible in this case, although \cite{Tsygankov2006} and \cite{Bykov2021} did report  hints of pulse profile changes.

Changes of the pulse profile shape around the CRSFs-energy or its harmonic have been reported for a few sources, e.g., V0332+53, 4U 0115+63 \citep{Tsygankov2006,Tsygankov2007}.
This phenomenon is probably linked to the specific radiation beaming near the cyclotron frequency \citep{Ferrigno2011,Schonherr2014}, or to the angular dependence of the CRSFs energy due to the Doppler shift produced by the bulk motion of the plasma towards the neutron star \citep{Mushtukov2015}, reflected in the angular dependence of the radiation intensity.

We attempted to investigate this phenomenon as a function of luminosity in 1A 0535+262. As seen in Figure \ref{pro_E_zong}, above 40 keV a sharp enhancement of the first peak and a phase jump of the second peak are clearly observed above the critical luminosity.
The energy dependence of the pulse profile of V0332+53 was also detected in relation to luminosity \citep{Tsygankov2006,Lutovinov2015}. At 3.4 $\times 10^{38}$ erg s$^{-1}$, the profile shapes are all similar across all energy bands (Left panel of Figure 9 in \cite{Tsygankov2006}) and becomes more complex above 7.3 $\times 10^{37}$ erg s$^{-1}$. In particular, the pulse profile changes from double peaks below 20 keV to single peak at 20--30 keV; notice that the CRSFs-energy of V0332+53 is about 28 keV.
The pulse profiles at 30--40 keV evolved again into a bimodal structure, but the dip and peak phase were clearly distinct to the low-energy (Right panel of Figure 9 in \cite{Tsygankov2006}).
4U 0115+63 also showed the energy dependence of pulse profile varies with luminosity. When the luminosity is above 7 $\times 10^{37}$ erg s$^{-1}$, the pulse profile exhibits a periodic phase lag in different energies. However, when the luminosity is below 7 $\times 10^{37}$ erg s$^{-1}$, the shape of pulse profiles barely vary with energy, except for a broadening of the pulse peak near the CRSFs-energy \citep{Tsygankov2007}. We conclude, therefore, that the complex evolution of the pulse profiles with energy and luminosity is not unique for 1A 0535+262. However, it has never been studied in such detail up to now.

The pulse fraction provides a quantitative description of the pulse intensity. The evolution of the pulse profile with energy dependence can leave a footprint in the pulse fraction-energy relation.
Indeed, the luminosity-dependence of the pulse fraction vs. energy relation was observed in V0332+53 and 4U 0115+63, corresponding to the luminosity range of the variation of the pulse profile with energy discussed above.
V0332+53 shows a hump-structure in the pulse fraction-energy relation at the fundamental CRSFs-energy \citep{Tsygankov2010}, while 4U 0115+63 with a dip-structure at the harmonic CRSFs-energy \citep{Tsygankov2007}.
However, this phenomenon has never been systematically investigated, and its dependent luminosity range has not been clearly reported.
From \textit{Insight-HXMT} observations of the 2020 giant outburst of 1A 0535+262, we for the first time identify the luminosity-dependence of dip-structure in the energy dependence of pulse fraction, which occurs only in the luminosity range from 4.8 $\times 10^{37}$ erg s$^{-1}$ to 1.0 $\times 10^{38}$ erg s$^{-1}$, covering both the sub-critical and super-critical stages of 1A 0535+262; the critical luminosity is $\sim$ 6.7 $\times 10^{37}$ erg s$^{-1}$ \citep{Kong2021}.
Comparing the results of V0332+53 and 4U 0115+63, the observations of 1A 0535+262 during the 2020 outburst make 1A 0535+262 the first source for which a limit on the luminosity of CRSFs-dependent structure in the pulse fraction-energy relation is observed. 
This luminosity range has not been clearly observed before, not only because of the scarcity of high cadence monitoring campaigns, but also because of the limited sample of sources observed entering the super-critical stage \citep{Staubert2019}. 

So far no model is available to interpret this phenomenon well.
A dramatic change in the pulse profile near the cyclotron line can be naturally accompanied with an increase or decrease of the local pulse fractions and we have been able to illustrate for the first time that this is indeed the case. Detailed interpretations of our results will require, however, the development of a sound model describing pulse profiles and spectra both below and above the critical luminosity, which is outside the scope of current work. Nonetheless, our results will be essential for testing such models.  

\section{Summary}

We report the brightest outburst in the history of 1A 0535+262 observations. Using high cadence and high statistics observations by \textit{Insight-HXMT}, we investigate the timing evolution properties of this source in detail. The main results are:

1. There is an energy dependence and a luminosity dependence of the pulse profile shape; the pulse profile exhibits significant changes around the luminosity corresponding to the expected transition between sub- and super-critical accretion regimes. On the other hand, the value of the critical luminosity estimated based on the observed CRSFs line energy-luminosity correlation \citep{Kong2021} is significantly higher than what could be derived based on the observed pulse profile evolution. This could indicate that the two estimates correspond to different transitions 
or to different phases of the same transition.

2. The RMS pulse fraction drops sharply near the CRSFs-energy. This local dip-structure at CRSFs-energy on the RMS pulse fraction-energy relation appears only near the critical luminosity, from 4.8 $\times 10^{37}$ erg s$^{-1}$ to 1.0 $\times 10^{38}$ erg s$^{-1}$. This result reinforces tight connection between observed spectra and pulse profiles.

This investigation presents an unprecedented study of the complete evolution of pulse profiles of 1A 0535+262. For the first time, the phenomenon of a local structure on the RMS pulse fraction-energy relation of 1A 0535+262 near the CRSFs-energy is revealed. Combined with other accreting pulsars, it is concluded that the local structure in pulse fraction-energy relation at CRSFs-energy may only appear near the critical luminosity. This luminosity-dependent relationship should be tested with more sources and more complete observations in the  future.

\acknowledgments
This work is supported by the National Key R\&D Program of China (2021YFA0718500), the National Natural Science Foundation of China under grants U2038101, U1838201, U1838202, U1938101, 12173103, U2031205, U1938103, 11733009, and Guangdong Major Project of Basic and Applied Basic Research (Grant No. 2019B030302001). This work made use of data from the \textit{Insight-HXMT} mission, a project funded by China National Space Administration (CNSA) and the Chinese Academy of Sciences (CAS). 
We thank the anonymous referee for valuable comments and suggestions.


\end{document}